\documentclass[manuscript,screen,review=false]{acmart}

\AtBeginDocument{%
  \providecommand\BibTeX{{%
    \normalfont B\kern-0.5em{\scshape i\kern-0.25em b}\kern-0.8em\TeX}}}

\setcopyright{acmcopyright}
\copyrightyear{2018}
\acmYear{2018}
\acmDOI{XXXXXXX.XXXXXXX}

\acmConference[ACM TAAS]{Special Issue on 20 Years of Organic Computing}{date}{location}
\acmPrice{15.00}
\acmISBN{978-1-4503-XXXX-X/18/06}

\usepackage[acronym,hyperfirst=false,nohypertypes={acronym},nonumberlist,nogroupskip]{glossaries}
\makeglossaries
\newacronym{cpes}{CPES}{cyber-physical energy system}
\newacronym{ict}{ICT}{information and communication technology}
\newacronym{ids}{IDS}{intrusion detection system}
\newacronym{it}{IT}{information technology}
\newacronym{ied}{IED}{intelligent electronic device}
\newacronym{isms}{ISMS}{information security management system}
\newacronym{oc}{OC}{organic computing}
\newacronym[firstplural=objects of investigation (OoIs)]{ooi}{OoI}{object of investigation}
\newacronym{ps}{PS}{power system}
\newacronym{psna}{PSNA}{power system network assessment}
\newacronym{psnatrust}{PSNA-Trust}{trust in power system network assessment}
\newacronym{cvc}{CVC}{coordinated voltage control}
\newacronym{rtu}{RTU}{remote terminal unit}
\newacronym{uml}{UML}{unified modeling language}
\newacronym{oltc}{OLTC}{on-load tap changer}
\newacronym{scada}{SCADA}{supervisory, control and data acquisition}

\usepackage{physics}

\begin{document}

\title{Applying Trust for Operational States of ICT-Enabled Power Grid Services}


\author{Michael Brand}
\authornote{Both authors contributed equally to this research.}
\email{michael.brand@offis.de}
\orcid{0000-0001-7037-2246}
\affiliation{%
  \institution{Energy Division, OFFIS -Institute for Information Technology}
  \city{Oldenburg}
  \country{Germany}
  \postcode{26121}
}
\author{Anand Narayan}
\authornotemark[1]
\email{anand.narayan@offis.de}
\orcid{0000-0003-3593-1780}
\author{Sebastian Lehnhoff}
\email{sebastian.lehnhoff@offis.de}
\orcid{0000-0003-2340-6807}
\affiliation{%
  \institution{Energy Division, OFFIS -Institute for Information Technology}
  \city{Oldenburg}
  \country{Germany}
  \postcode{26121}
}
\affiliation{%
  \institution{Carl von Ossietzky University of Oldenburg}
  \city{Oldenburg}
  \country{Germany}
  \postcode{26129}
}

\renewcommand{\shortauthors}{Brand and Narayan et al.}

\begin{abstract}
Digitalization enables the automation required to operate modern cyber-physical energy systems (CPESs), leading to a shift from hierarchical to organic systems. However, digitalization increases the number of factors affecting the state of a CPES (e.g., software bugs and cyber threats). In addition to established factors like functional correctness, others like security become relevant but are yet to be integrated into an operational viewpoint, i.e. a holistic perspective on the system state. Trust in organic computing is an approach to gain a holistic view of the state of systems. It consists of several facets (e.g., functional correctness, security, and reliability), which can be used to assess the state of CPES. Therefore, a trust assessment on all levels can contribute to a coherent state assessment. This paper focuses on the trust in ICT-enabled grid services in a CPES. These are essential for operating the CPES, and their performance relies on various data aspects like availability, timeliness, and correctness. This paper proposes to assess the trust in involved components and data to estimate data correctness, which is crucial for grid services. The assessment is presented considering two exemplary grid services, namely state estimation and coordinated voltage control. Furthermore, the interpretation of different trust facets is also discussed.

\end{abstract}



\maketitle

\section{Introduction}\label{sec:introduction}

\subsection{Motivation}\label{sec_motivation}

Modern power systems have increased uncertainties due to the penetration of distributed energy resources. The safe and reliable operation of such systems requires improved observability and controllability. This is achieved by the deep interconnection with \gls{ict} systems, giving rise to \glspl{cpes}~\cite{Yu2016}. A promising approach for operating such complex systems with system-wide \gls{ict}, a large number of actors, and a high degree of automation is to shift from a hierarchical system, operated mainly in a centralized manner by humans, to a distributed one, mainly operated autonomously but with several actors interacting with each other as well as with humans. Consequentially, such \glspl{cpes} can be regarded as organic computing systems~\cite{niesse2016controlled, loeser2022vision} as they continuously and dynamically adapt to exogenous and endogenous changes on various time scales -- from significant changes in generation technologies driven by external regulatory and political factors to unforeseen demand fluctuations due to stochastic human behaviour. To cope with these dynamics, the system is characterized by self-* properties, e.g., self-organization of battery storage swarms \cite{tiemann2022operational}, self-configuration of SCADA and protection systems \cite{velasquez2019flexible}, self-explanation, and context awareness~\cite{brand2019framework}.

Owing to the geographical size and many actors, \glspl{cpes} are among the biggest and most complex human-made organic computing systems. However, the \gls{ict} penetration has further increased the overall system complexity, leading to new risks such as software failures, malfunctions, and cyberattacks~\cite{tondel2018interdependencies}. This, in turn, increases the number of factors affecting the state and health of a CPES. In addition to established factors like functional correctness and safety, other factors like security, credibility, and usability must be considered. Although these aspects may already be considered in certain subsystems (e.g., safety for power lines and security for communication networks), they are yet to be integrated and used to operate the whole CPES, i.e., a holistic state and health of CPESs considering the individual subsystems. 

Following the idea of the transition of \glspl{cpes} towards organic computing systems, trust from the domain of \gls{oc} (\gls{oc}-Trust)~\cite{Steghofer2010}, used for assessing trust between autonomous actors and technical subsystems, is a promising approach for a holistic state and health assessment. Here, trust is defined as a context-dependent and multifaceted sense of a technical entity with respect to its functional correctness, safety, security, reliability, credibility, and usability. All of these facets can be directly mapped onto the state and health of complex \glspl{cpes}. The assessment of trust on all levels of a \gls{cpes} can contribute to a coherent state and health assessment, which may then be interpreted by technical subsystems or used by operators in their decision-making.

Assessing the trust of \glspl{cpes} implies assessing the trust of its constituting (autonomous) subsystems, components, and services, which is challenging as they face a broad and diverse threat landscape~\cite{tondel2018interdependencies}. In addition to disturbances in power systems, disturbances in \gls{ict} systems can also impact the interconnected power system via the grid services. The 2003 North American blackout resulted from a software bug in the state estimation service, which gave incorrect situational awareness to the system operators, thereby hiding certain power line failures \cite{NERC2004}. The 2015 and 2017 Ukraine blackouts were caused by hackers taking over control rooms and shutting down vital grid services \cite{Wu2017}. This illustrates the vulnerability of \glspl{cpes} to cyberattacks. Other examples of \gls{ict} disturbances include sensor failures affecting observability, controller failures affecting controllability, and communication network failures affecting data transfer, causing data to be either delayed or lost \cite{narayan2019first}. It is evident that \gls{ict} disturbances impact the performance of grid services, which in turn impacts the operation of the power system. Therefore, these ICT-enabled grid services can be regarded as entities, for which trust aspects like functional correctness, security, reliability, and credibility must be assessed.

\begin{figure}[htb]
\centering
\includegraphics[width=.85\textwidth]{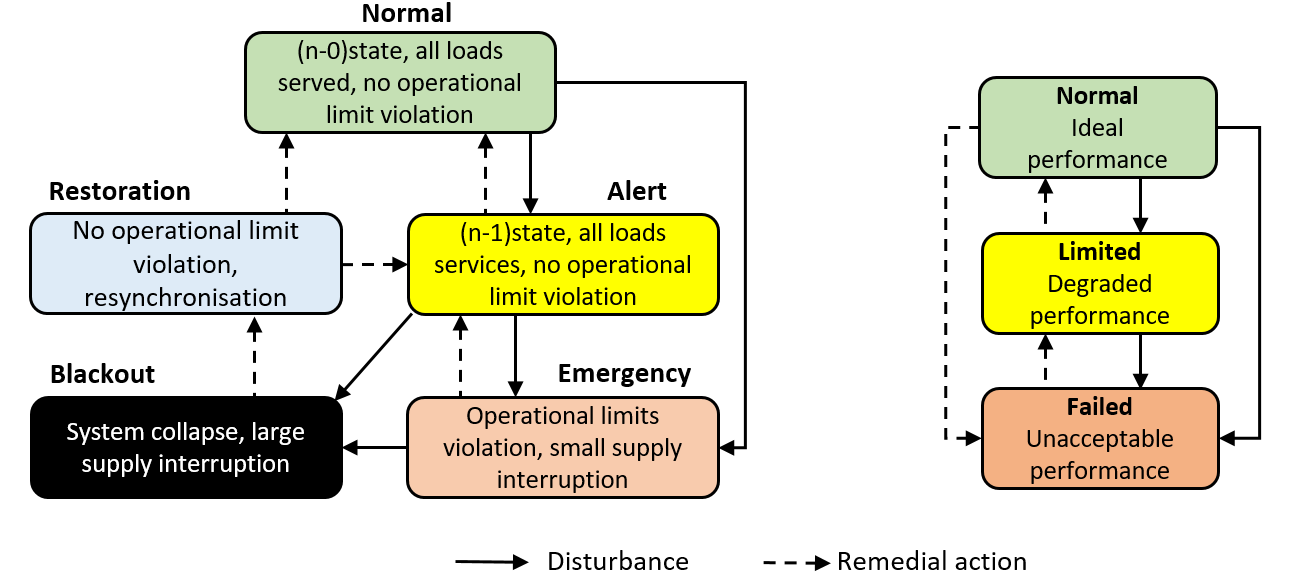}
\caption{Operational states and transitions of Power System (left) and ICT-enabled Grid Service (right)~\cite{klaes2020state}}
\label{fig:example_states}
\end{figure} 

The operational state classification solely based on electrotechnical parameters (e.g., voltage, frequency, currents) is the prominent state-of-the-art among power system operators to assess the current state (or performance) of a power system~\cite{EUCom1485_SysOp, fink1978operating}. The operation of a power system can be classified into one of five different states: \textit{normal}, \textit{alert}, \textit{emergency}, \textit{blackout}, and \textit{restorative}. Disturbances in the system can cause the state to degrade and suitable grid services (referred to as remedial actions) can be triggered to improve the state again. However, as already mentioned, certain disturbances can also impact the grid services, causing them (in the worst case) to fail \cite{narayan2019first}. To address this, the authors of \cite{klaes2020state,haack2022hybrid} propose an operational state classification for each \gls{ict}-enabled grid service. Based on three properties of the \gls{ict} system, i.e., availability of components and data, timeliness of data transfer, and data correctness, the state of a grid service can be classified into \textit{normal}, \textit{limited}, or \textit{failed}. These properties are regarded as the requirements of the grid services. Figure \ref{fig:example_states} shows the operational states of the power grid as well as of the ICT-enabled grid services, along with possible state transitions~\cite{klaes2020state}. The remedial actions in the operational state classification of a power system (left part) are performed via the ICT-enabled grid services and the remedial actions of grid services (right part) would be actions in the ICT network, such as rerouting traffic or traffic shaping~\cite{zhao2017off}. The states of grid services will be further elaborated in Section \ref{sec_classification}. The authors of~\cite{klaes2020state} present only the theoretical foundations for these states without concrete use cases. These states are formalized in \cite{haack2022hybrid} and their benefits were demonstrated using simulations in~\cite{narayan2021analyzing}. However, the impact of data correctness on the state of ICT-enabled grid services is, however, not investigated because the ICT disturbances impacting the state are not considered.

As mentioned earlier, cyberattacks are disturbances that can impact all three aforementioned properties. A denial-of-service attack can render a component unable to communicate (availability) or communicate with unaccepted timeliness~\cite{Wang2018Cyber}, whereas a false data injection attack can impact the data correctness by injecting manipulated data~\cite{Liu2011}. In this regard, there exists a fundamental difference between the aforementioned availability and timeliness on one hand and data correctness on the other. The first two properties can be measured, e.g., by pinging (or heartbeat) and comparing timestamps, respectively~\cite{kruger2020real}. However, data correctness cannot be directly measured due to the absence of ground truth (i.e., real/actual value).


Traditionally, data correctness in a power system refers to the correctness of measurements. State estimation, in combination with bad data detection, is used in this regard and is the base for almost all other services~\cite{narayan2019towards}. However, this requires redundant measurements and assumes measurement errors to be randomly distributed~\cite{Abur2004}. While the former is not applicable for distribution grids (especially on the lower voltages)~\cite{dehghanpour2018survey}, the latter does not hold for all ICT disturbances, e.g., for coordinated false data injection attacks~\cite{Liu2011}. Therefore, data correctness is challenging to determine, especially considering the rising cyber threats in \glspl{cpes}. Unlike other disturbances, such as link and sensor failures, the diverse nature of cyberattacks makes it challenging to investigate their impact on the \gls{ict} system as well as the interconnected power system~\cite{brand2019trust}. Furthermore, since \gls{ict} systems span over large geographical areas (similar to power systems), data transfer from source to destination typically passes through several intermediate nodes (or hops), all of which are potential entry points for attackers. An intelligent attacker can make it challenging (or even impossible) for destination nodes to determine, whether the received data is correct, especially without the ground truth. Additionally, as shown in~\cite{narayan2019first}, malfunctions and incorrect operation of components (e.g., sensors) can also impact data correctness.

\subsection{Related Work}\label{sec_SOA}

Addressing issues regarding the correctness of process data has been the focus of several studies. They can be categorized into field measures (e.g.,~\cite{dan2010stealth, pei2020pmu}), improved bad data detection (e.g.,~\cite{Glazunova2014, wang2020detection}), and trust-based measures (e.g.,~\cite{Liu2018Secure, brand2019trust}). Field measures aim to improve cyber security and accuracy, e.g., by placing different types of sensors, but not by estimating data correctness. Research on improved bad data detection is specialized (e.g., against coordinated false data injection attacks). Both measures in the field and improved bad data detection only tackle a subset of threats impacting the correctness of measurements or they rely on certain invalid assumptions (e.g., measurement redundancy). Considering various types of power systems, grid services, and threats to the correctness of measurements, a more holistic approach capable of integrating existing approaches is needed. Trust-based approaches can provide a suitable framework to estimate data correctness, but the interpretation of the term trust lacks uniformity. In~\cite{Liu2018Secure}, for example, field devices (referred to as agents) derive trust from the deviation of the estimated measurements of neighbouring devices based on measurements received from the neighbouring device. A disadvantage of such univariate trust-based measures is that the trust in the measurements of the other device relies only on one piece of information, in this case, the deviation from the expected data.

The authors of~\cite{brand2019trust, brand2020trust, brand2021trust} adapt \gls{oc}-Trust to a multifaceted trust model for power systems, referred to as \gls{psnatrust}. Here, trust is defined as a "context-dependent, and multivariate sense about an entity with respect to its functional correctness, safety, security, reliability, credibility, and usability". Additionally, a methodology to assess or estimate the trust in measurements and state variables is presented in~\cite{brand2020trust, brand2021trust}. Due to the absence of ground truth, data correctness can often only be estimated (i.e., cannot be directly measured), which makes \gls{psnatrust} and its assessment methodology a promising approach.

There also exist other trust approaches in other domains like the so-called ABI model (ability, benevolence, integrity) in the field of organizational trust~\cite{mayer1995integrative} or the concept of data veracity in the field of big data~\cite{assiri2020methods, krotofil2015process}. However, \gls{oc}-Trust focuses on trust in and between autonomous software agents in complex systems and provides for this specific environment, a more specific (compared to data veracity) and fine-granular (compared to the ABI model) grouping of factors affecting trust. In addition, there already exist applications of OC-Trust in the domain of power systems~\cite{Anders2011Patterns, Rosinger2013Threat, brand2020trust}.

\subsection{Contribution}\label{sec_contribution}
This paper aims to combine a definition of states of \gls{ict}-enabled grid services~\cite{klaes2020state,narayan2021analyzing} with a trust assessment model~\cite{brand2019trust, brand2020trust, brand2021trust} by using trust to estimate data correctness. Overall, this paper highlights the need for a state classification for ICT-enabled grid services and serves as a step towards the system-wide application of trust to assess the state or health of a CPES. The concrete contributions of this paper are:

\begin{itemize}
    \item an interpretation of the trust facets for estimating data correctness considering various ICT disturbances,
    \item the application of trust as a measure of data correctness for \gls{ict}-enabled grid services, and
    \item two use cases of grid services, namely state estimation and \gls{cvc}, along with a concrete approach to derive the state of these services using trust. 
\end{itemize}

Since this paper is a conceptual work, a concrete implementation is out of its scope. The remainder of this paper is structured as follows. Section~\ref{sec_classification} introduces the operational state classification for \gls{ict}-enabled grid services and the properties used for the classification. The application of trust as a measure for data correctness is explained in Section~\ref{sec:trust}, which also includes an overview of \gls{psnatrust} and the interpretation of the trust facets (functional correctness, safety, security, reliability, credibility, and usability) in the context of the targeted application. In Section~\ref{sec:use_cases}, the operational state classification considering \gls{psnatrust} as a measure of data correctness is presented based on two exemplary grid services, namely state estimation and \gls{cvc}. Section~\ref{subsec:discussion} discusses the proposed approach, limitations, and future work. Finally, Section~\ref{sec:conclusion} concludes the paper.

\section{State Classification of ICT-enabled Grid Services}\label{sec_classification}

A typical ICT system in a CPES encompasses hardware and software components for data acquisition, control (or actuation), computation, and data transfer. In the context of this paper, the ICT system includes power system field devices such as sensors and controllers, communication network devices such as routers and switches, and servers for computation located in the control room. The role of the ICT system is to enable the grid services by fulfilling their aforementioned requirements of suitable data/components, namely availability, timeliness, and correctness. Consequentially, ICT disturbances can impact the performance of these grid services. Based on~\cite{klaes2020state}, the states of grid services, which reflect their performance, can be classified as follows: 

\begin{itemize}
    \item \textbf{Normal State:} In this state, the grid service is fully functional and can be used as intended. Coordinated decision-making and control are possible in this state. A grid service is said to be in \textit{normal} state if no disturbance has occurred or if the occurred disturbances have been absorbed by the robustness of the ICT system (e.g., redundant components). 
        
    \item \textbf{Limited State:} This state indicates partial performance degradation of the grid service and the service should be used with caution. This state is typically characterized by disturbed communication, which limits the coordination among various actors. A grid service is in its \textit{limited} state if certain disturbances have negatively impacted the service, causing it to resort to fallback mechanisms (e.g., using local instead of wide-area measurements).  
        
    \item \textbf{Failed State:} In this state, the service is no longer functional or yields grossly incorrect results. Suitable actions should immediately be taken to restore the functionality of the service. A service is said to be in \textit{failed} state when disturbances have impacted critical components beyond the scope of the fallback mechanisms. 
\end{itemize}

These states are further elaborated in Section~\ref{sec:use_cases}. As shown in~\cite{klaes2020state, haack2022hybrid}, the state of a grid service can be assessed using the following three properties.

\begin{itemize}
    \item \textbf{Availability} indicates the functionality of a component at a specific time instant. Each grid service requires certain ICT components (e.g., sensors, routers, and servers) to be available. Depending on its design, the service can still be in the \textit{normal} state despite the unavailability of certain non-critical components. However, disturbances causing the unavailability of critical components will lead to a performance degradation of the grid service.  
    
    \item \textbf{Timeliness} indicates, whether a grid service can provide a timely reaction. A grid service should be able to receive field measurements, take decisions, and then send control signals to the actuators in a timely manner. In the context of communication networks, timeliness refers to latency, which is a quality-of-service aspect that captures the speed of data transfer (e.g., processing and transmission delays in routers).

    \item \textbf{Data Correctness} is the closeness of the data (measurement or control signal) to its ground truth. Data correctness heavily influences the decisions made by the grid services based on this data.   
\end{itemize}
   

   These properties have service-specific thresholds, which are further elaborated in Sections~\ref{subsec:stateestimation} and~\ref{subsec:voltagecontrol}. As mentioned in Section~\ref{sec:introduction}, measuring availability and latency are well-researched topics. Availability of components can be determined by monitoring their heartbeats as shown in \cite{kruger2020real}. Latency could be measured using network monitoring tools such as CheckMK \cite{brand2019framework}. However, the same is not true for data correctness. Measuring data correctness is challenging as the ground truth is typically not known. To address this, the concept of PSNA-Trust is considered in the operational state classification of ICT-enabled grid services.

\section{Trust as a Measure of Data Correctness}
\label{sec:trust}
Trust in the context of \gls{psna}~\cite{brand2019trust,brand2020trust} is defined as a "[...] context-dependent, and multivariate sense about an entity with respect to its functional correctness, safety, security, reliability, credibility, and usability". While functional correctness is a so-called trust facet, data correctness, in the scope of this paper, is a property of an ICT-enabled grid state (cf. Section~\ref{sec_classification}). Since data correctness is challenging to assess, this paper proposes to assess the trust in components, data, and services and to use the assessed trust as hints for data correctness. If, for example, a component has issues with functional correctness, the trust in the component and data from that component should be decreased, which should be reflected in the state of the service accordingly. Here, trust in process data could be estimated based on static information about the involved components (e.g., from an \gls{isms} for the security facet), live information from monitoring systems (e.g., from an \gls{it} health monitoring system for the functional correctness facet), and experience~\cite{brand2020trust, brand2019framework, brand2021trust}.


A model for trust in the context of power systems called \gls{psnatrust} and the estimation of trust of components and data is presented in Subsection~\ref{subsec:psna_trust}. In Subsection~\ref{subsec:trust_interpretation}, the role of the individual trust facets for data correctness is discussed. This is relevant as not all facets have the same influence on data correctness. An implementation of \gls{psnatrust} with numerical calculations for the use cases is out of this paper's scope. However, a demonstration of the application of the trust model for a state estimation service is described in~\cite{Brand2021Dach}\footnote{Demo video: \url{https://youtu.be/3hwi49sfllQ}}.

\subsection{The Model of PSNA-Trust}
\label{subsec:psna_trust}

\begin{figure}[htb]
\centering
\includegraphics[width=.96\textwidth]{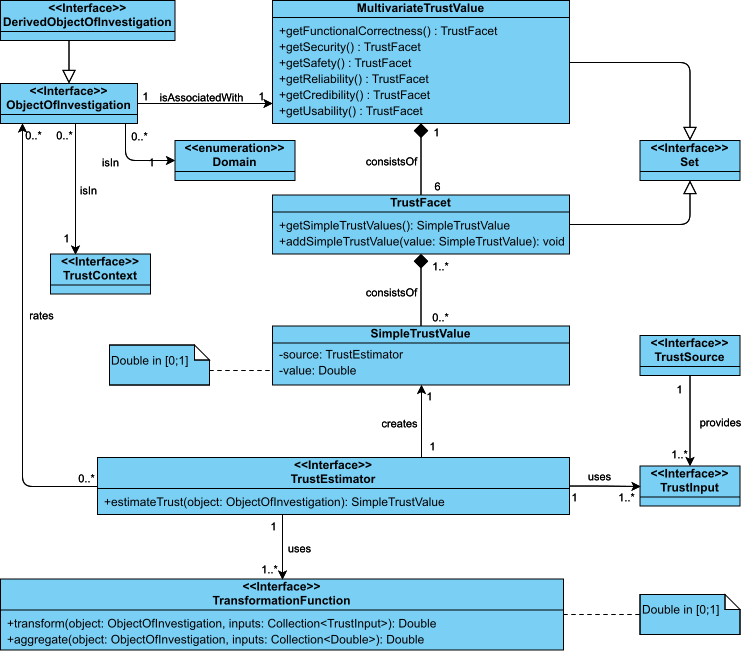}
\caption{The model of \gls{psnatrust} as an \gls{uml} class diagram.}
\label{fig:psnatrust}
\end{figure}

Figure~\ref{fig:psnatrust} shows the model of \gls{psnatrust} as an \gls{uml} class diagram. The model combines the representation of trust with its facets, i.e., functional correctness, safety, security, reliability, credibility, and usability, and an approach for its estimation. In the top left, there are the \glspl{ooi} -- the entities (e.g., components) for which trust is to be estimated. \Glspl{ooi} can be clustered into domains such as power and \gls{ict} systems. Another part of \gls{psnatrust} is the derived \glspl{ooi}, the trust in which can only be estimated indirectly using the trust in other \glspl{ooi}. For example, in the acquisition of measurements in a power system, the measurements themselves are not monitored but the components involved in the acquisition process, such as sensors and routers. Hence, the trust in such measurements (derived \glspl{ooi}) is derived using the trust in the involved components (\glspl{ooi}).

An important aspect of \gls{psnatrust} is that there can be various information about specific trust aspects within a facet. A security aspect can be derived from the information from an \gls{isms} as well as from an \gls{ids}. \gls{psnatrust} considers this by allowing different trust estimators with different trust sources in parallel and by defining so-called simple trust values~\cite{brand2021trust}. A simple trust value is a pair consisting of a trust estimator and a trust probability, i.e. an estimation of the trust in the \gls{ooi} from the perspective of the respective trust estimator. These are also shown in Figure~\ref{fig:psnatrust}. Trust sources on the right side provide trust inputs that are used by trust estimators to estimate a trust probability, for example, alerts (trust inputs) from an \gls{ids} (trust source). To do so, the trust estimators use transformation functions that transform the trust inputs for a given \gls{ooi} in a certain context to a trust probability. An example of a trust context is the temporal validity of the \gls{ooi}. Trust inputs should only be used for a certain \gls{ooi} if they match in time. Furthermore, transformation functions can also aggregate different trust probabilities, which is needed for derived \glspl{ooi}. A trust estimator can, in the first step, calculate the trust probability for all involved (direct) \glspl{ooi} and, in the next step, aggregate the results to derive a trust probability for the derived \gls{ooi}.

Irrespective of the number of trust estimators and the concrete transformation functions, the result is a set of simple trust values, which are mapped to at least one trust facet. Therefore, a trust facet is a set of simple trust values and can be empty if there is no trust estimator contributing to that facet. Accordingly, a multivariate trust value (top in Figure~\ref{fig:psnatrust}) is a set of different trust facets and, therefore, a set of sets of simple trust values. Based on the definition of \gls{psnatrust}, a multivariate trust value is a 6-tuple containing the facets functional correctness, safety, security, reliability, credibility, and usability.

\begin{figure}[htb]
\centering
\includegraphics[width=.96\textwidth]{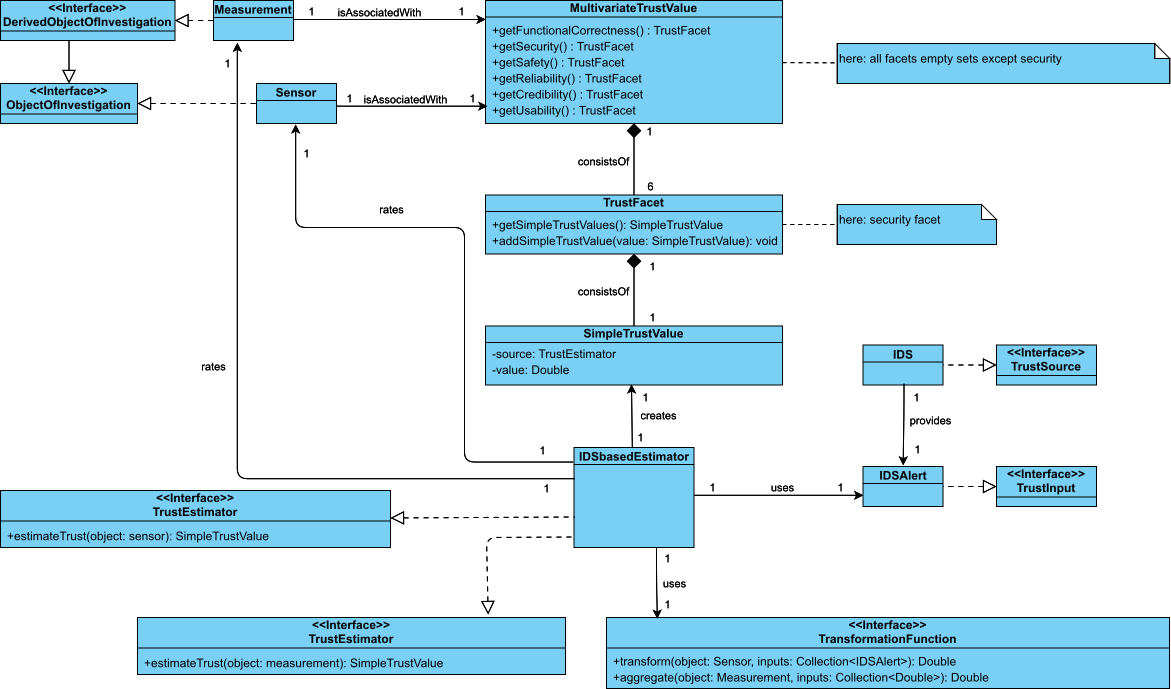}
\caption{Examplary instantiation of \gls{psnatrust} as an \gls{uml} class diagram.}
\label{fig:psnatrust_example}
\end{figure}

Figure~\ref{fig:psnatrust_example} shows a minimal exemplary instantiation of the model with sensors as \glspl{ooi} and measurements as derived \glspl{ooi}. A trust estimator, called \texttt{IDSbasedEstimator}, uses alerts from an \gls{ids} to estimate the trust (specifically, the security) in sensors as well as in measurements. To do so, it uses a transformation function that is capable of transforming \gls{ids} alerts to a trust probability for a sensor and of aggregating trust probabilities for sensors (in this example, only one sensor is considered) to a trust probability for a measurement. The resulting simple trust values contribute to the security facet of the sensor or the measurement, respectively.

The described model is for the estimation of trust in components and data. However, for the grid service property of data correctness, the estimation of the trust in service outputs is required, which depends on the trust in the inputs and the service core (algorithm). This dependency is specific for each grid service as it depends on the service algorithm and the dependency between the inputs and outputs. For the state estimation service in a power system, an estimation of the trust in the outputs (state variables) based on the trust in the inputs (field measurements) is proposed in~\cite{brand2021trust}. The result is a multivariate trust value for each state variable, where the information about different facets and simple trust values is kept intact. However, to estimate data correctness, it is necessary to aggregate the multivariate trust values, first, to a single multivariate trust value for the whole service output and, second, to a single trust probability. Furthermore, not all trust facets are feasible or relevant to estimate the data correctness. This is discussed in the following subsection. 

\subsection{An Interpretation of the Trust Facets}
\label{subsec:trust_interpretation}
While the PSNA-Trust model has already been used (e.g. in~\cite{brand2020trust, brand2021trust}), an interpretation of the trust facets has only been discussed with a low fidelity in~\cite{brand2019trust}. Therefore, this subsection provides a more elaborated interpretation of the trust facets with the goal of using trust to estimate data correctness. For \gls{psnatrust}, the trust facets are defined as follows~\cite{brand2019trust}, which are adapted from \gls{oc}-Trust~\cite{Steghofer2010}.

\begin{itemize}
    \item \textit{Functional correctness} is ''the quality of a system to adhere to its functional specification under the condition that no unexpected disturbances occur in the system’s environment''.
    \item \textit{Safety} is ``the quality of a system to be free of the possibility to enter a state or to create an output that may impose harm to its users, the system itself or parts of it, or to its environment''.
    \item \textit{Security} is ``the absence of possibilities to defect the system in ways that disclose private information, change or delete data without authorization, or to unlawfully assume the authority to act on behalf of others in the system''.
    \item \textit{Reliability} is ``the quality of a system to remain available even under disturbances or partial failure for a specified period of time as measured quantitatively by means of guaranteed availability, mean-time between failures, or stochastically defined performance guarantees''.
    \item \textit{Credibility} is ``the belief in the ability and willingness of a cooperation partner to participate in an interaction in a desirable manner. Also, the ability of a system to communicate with a user consistently and transparently''.
    \item \textit{Usability} is ``the quality of a system to provide an interface to the user that can be used efficiently, effectively and satisfactorily that, in particular, incorporates consideration of user control, transparency and privacy''.
\end{itemize}

As mentioned in the previous section, all of the six facets do not necessarily contribute to an estimation of data correctness. Reduced reliability, for example, is of high relevance for (predictive) maintenance of components but not for data correctness at the point in time the component delivers the data. Therefore, it is proposed to build context-dependent clusters of facets. In the scope of data correctness, a cluster for facets that directly contribute to the correctness of components or data, referred to as \textit{data correctness cluster}, is proposed. This contains the following facets:

\begin{itemize}
    \item Reduced \textit{functional correctness} of a component, data, or service may indicate that the respective entity is not working correctly or that the data is not correct, respectively. A memory load at maximum, for example, can lead to data loss and, therefore, wrong results.
    
    \item Reduced \textit{security} in terms of reduced integrity of a component, data, or service, on the one hand, may indicate that the respective entity is manipulated, hints for a manipulation exist, or that it is at least vulnerable for manipulations, i.e. data correctness cannot be guaranteed. A reduced availability of information (as part of information security), on the other hand, may indicate that information/data from a component or service is not available. This can also reduce data correctness. Alerts from an \gls{ids}, for example, can be the result of a data manipulation or denial-of-service attacks.
    
    \item Reduced \textit{credibility} may indicate that a component, data source, or service cannot or does not want to work correctly or provide correct data. A third-party system, for example, cannot be monitored to the same extent as its own system, and, therefore, its credibility cannot be guaranteed.
    
    \item Reduced \textit{usability} of a component, data source, or service may have caused wrong handling by a human, thereby impacting the data correctness. This plays a role only if entities are involved that are controlled by humans. If, for example, it is not clear in which unit a certain value has to be entered in a user interface, the value might be wrong because of bad usability. Another example of usability pertains to the cognitive load of the human operator. Due to the large number of digital actors, CPESs have large volumes of data, which can potentially overwhelm the human operators resulting in wrong decisions. Consequentially, systems should be designed considering the strengths of human perception to enable fast and correct usage by the human operator~\cite{saager2022konect}.  
\end{itemize}

The described interpretation of the facets with the aim of using trust to estimate data correctness considering various ICT disturbances is the first contribution of this paper (cf. Subsec.\,\ref{sec_contribution}). In the following section, this is used in the state assessment of the ICT-enabled grid services.

\section{Operational States of ICT-enabled Grid Services}\label{sec:use_cases}

\begin{figure}[htb]
    \centering
    \includegraphics[width=.75\textwidth]{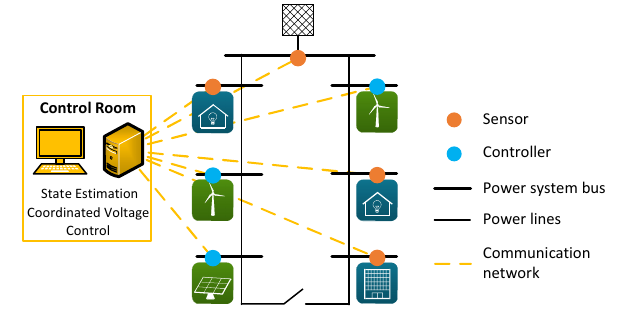}
    \caption{Exemplary grid showing state estimation and coordinated voltage control services.}
    \label{fig:example_grid}
\end{figure}

In this section, the feasibility of the proposed state classification is demonstrated. An exemplary \gls{cpes}, the two ICT-enabled grid services of state estimation and \glsreset{cvc}\gls{cvc}, and their operational states considering PSNA-Trust are presented in detail. Figure \ref{fig:example_grid} shows the exemplary \gls{cpes}. The thicker solid lines represent the power system buses and the thinner solid lines represent the electric power lines. These are analogous to nodes and edges of a graph. Measurements from different parts of the grid are gathered from ICT-based sensors. They are transmitted via the communication network (yellow dashed lines) to the control room, where the servers for processing are located. In this context, the control room represents the system operator. In a typical power system, the state estimation service runs centrally on a control room server \cite{Abur2004}. 
The controllable elements in the grid (in this example, a photovoltaic panel and wind turbines) can be controlled via ICT-based controllers, which have two operational modes: remote and local. In the remote mode, a central coordinated control strategy is followed, where control signals (referred to as setpoints) are sent from the control room to the controllers via the communication network. The \gls{cvc} service typically involves controlling different power system parameters based on the state estimation results to remedy voltage violations. Examples of \gls{cvc} implementations, which depend on the state estimation service, can be found in \cite{salih2015coordinated,hird2004network,viawan2008coordinated}. Without setpoints from the control room, the controllers shift to the local mode of operation as a fall-back measure. In this mode, the controllers act solely based on local measurements and are, therefore, uncoordinated. 

In the following subsections, the state classification for the two grid services, namely, state estimation and \gls{cvc} are proposed. Note that many possibilities exist for the state classification of ICT-enabled grid services. The presented ones are just options to emphasize the need for such a state classification.

\subsection{Operational States of State Estimation}\label{subsec:stateestimation}

State estimation is one of the important ICT-enabled services used to monitor the power system in real-time by estimating the two system state variables, i.e. voltage magnitude and angle at all buses~\cite{Abur2004}. This is done based on measurements gathered from sensors in the power system field as well as using a topological model of the power system. Typical field measurements are currents, active, and reactive power at different parts of the power system. State estimation provides situational awareness to the grid operator, based on which suitable control actions could be taken to improve the power system state (cf. Fig\,\ref{fig:example_states}). 

Weighted least squares is one of the most common state estimation algorithms~\cite{Abur2004, dehghanpour2018survey}. The necessary and sufficient condition for the solvability of this algorithm is $rank(H) = n_{sv}$, where $H$ is a function that relates the available field measurements to the number of state variables ~\cite{lukomski2008methods}. In addition to the availability of a server for running the state estimation algorithm, a successful run of this service requires sufficient measurements to satisfy this criterion. State estimation is performed at fixed time intervals with the measurements available at the control room, which imposes a timeliness (or latency) constraint on the field measurements to reach the control room. Typical latency requirements for state estimation can be found in~\cite{Kansal2012, Kuzlu2014}. It can vary depending on the design of this service and is also a design specification for the communication network in the \gls{cpes}. 

Furthermore, since the weighted least squares algorithm is deterministic, the correctness of state estimation results, i.e., the state variables, largely depends on the correctness of the input field measurements. This implies that only the measurements with a certain level of data correctness should be considered by the state estimation algorithm. A typical state estimation service is equipped only with a bad data detector, as discussed in Section~\ref{sec:introduction}, which is insufficient for estimating data correctness, especially considering disturbances that are not randomly distributed, e.g., coordinated cyber attacks.

Failures of sensors or congestions in the \gls{ict} system may either result in a complete loss or a delayed arrival of certain measurements at the control room. The correctness of measurement data received at the control room can be impacted by ICT disturbances. In case of temporal unavailability or incorrectness of field measurements, they can be replaced by suitable pseudo-measurements, which are typically calculated based on historical measurements~\cite{dehghanpour2018survey}. This is a fallback measure that, on the one hand, ensures solvability but, on the other hand, decreases the data correctness of the state estimation results as pseudo-measurements may not capture recent events in the \gls{cpes}, i.e. can be outdated. Based on these criteria, the state classification (cf. Section~\ref{sec_classification}) and the trust assessment (cf. Section~\ref{sec:trust}), the following operational states of the state estimation service reflecting its performance can be derived:

\begin{itemize}
    \item \textbf{Normal State}: The state estimation service is in the normal state if all state variables can be estimated properly, i.e., the solvability is achieved only with field measurements (no pseudo-measurements) and all state variables can be trusted in terms of the facets of the data correctness cluster.
    
    
    \item \textbf{Limited State}: The state estimation service is in the limited state if pseudo-measurements were needed to either achieve solvability or achieve trust in a facet of the data correctness cluster. Due to pseudo-measurements, the resulting state variables should be used with caution. Appropriate actions are also demanded to improve the state of the grid service even though the service has not failed. 
    
    \item \textbf{Failed State}: The state estimation service is in the failed state if it cannot be run at all, the solvability cannot be achieved even with pseudo-measurements, or if the resulting state variables cannot be trusted in terms of at least one facet of the data correctness cluster, even after using pseudo-measurements. In this state, the state variables are either unavailable or cannot be trusted. This is henceforth referred to as \textit{data correctness issue}.  
\end{itemize}


\begin{figure}[htbp!]
    \centering
    \includegraphics[width=.9\textwidth]{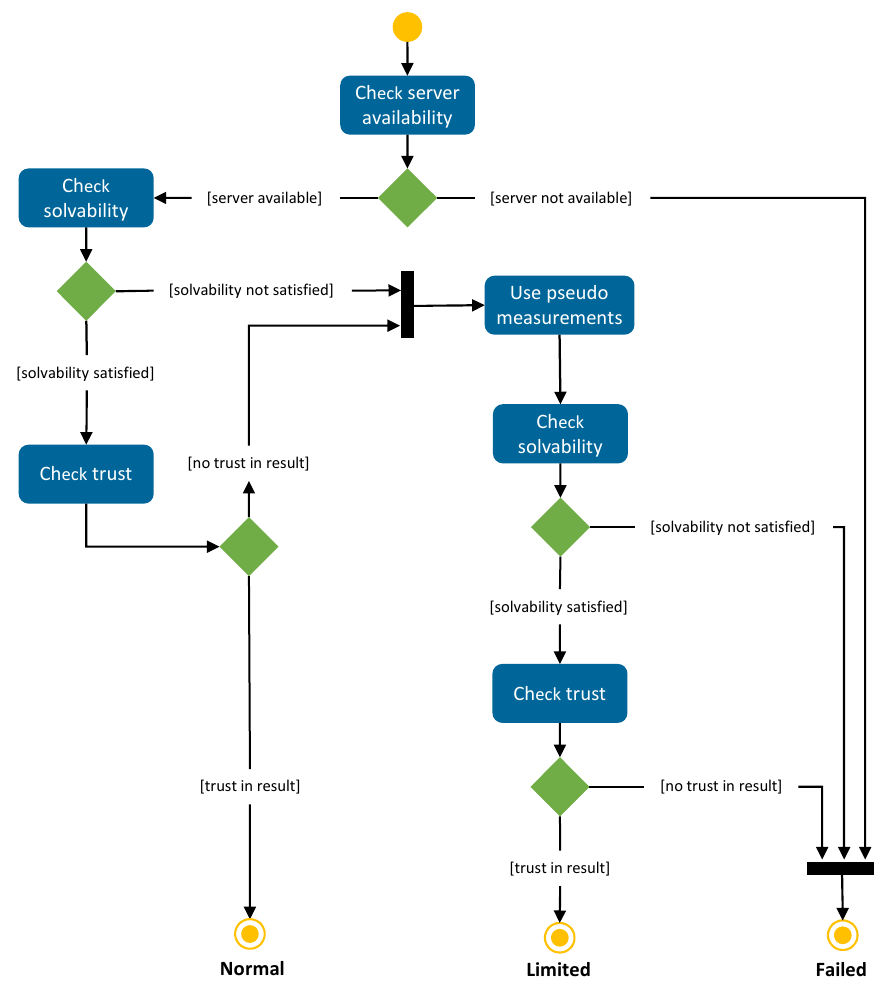}
    \caption{Activity Diagram showing states of State Estimation Service.}
    \label{fig:SE_activity}
\end{figure}

Figure~\ref{fig:SE_activity} shows an activity diagram for assessing the state of a state estimation service. First, the availability of the server, which hosts the state estimation algorithm, is checked. This is trivial and the unavailability of a suitable server results in the failed state. Typical \glspl{cpes} have redundant servers, where a backup server takes over in the case of primary server failure~\cite{kruger2020real}. If at least one server is available, the solvability is checked based on the measurements available within the specified time interval. If this criterion is satisfied, the state estimation algorithm is run, and the state variables are calculated. The trust in these state variables (cf.~\cite{brand2021trust}) regarding data correctness issues is then checked. Here, thresholds at which state estimation results can be trusted must be defined. Furthermore, the multivariate trust values have to be aggregated to calculate a trust probability that can then be compared with this threshold. Depending on the design of state estimation, this could be done in several ways such as calculating the minimum or (weighted) average~\cite{brand2021trust}.

If the solvability condition is satisfied and if there are no data correctness issues, the service is in the normal state. 
If the solvability is violated or if there is a data correctness issue in the state variables, suitable pseudo-measurements are integrated and the solvability as well as the trust in the state variables are successively re-evaluated. Since pseudo-measurements are derived from historical data, there exists a certain design-specific threshold for their maximum number that could be used for a state estimation run~\cite{dehghanpour2018survey}. It is possible to increase the use of pseudo-measurements and perform the two aforementioned checks iteratively until either both checks are satisfied, or the threshold for the maximum pseudo-measurements is reached (without satisfying the checks). The former (both checks satisfied) would indicate that the service is in the limited state, while the latter would indicate a failed state.

\begin{align}
    \begin{split}\label{eq:se_failed}
    S_{SE} = \text{failed} ~\Leftrightarrow~ & \neg\Phi_{serv} \lor ((rank(H_{\va{z}}) \neq n_{sv} ~\lor~ t_c(\va{x},\va{z}) < t_{c,threshold})\\ & \land ~ (rank(H_{\va{z_p}}) \neq n_{sv} ~\lor~ t_c(\va{x},\va{z_p}) < t_{c,threshold}))
    \end{split}\\
    \begin{split}\label{eq:se_ltd}
    S_{SE} = \text{limited} ~\Leftrightarrow~ & \Phi_{serv} ~\land~ (rank(H_{\va{z}}) \neq n_{sv} ~\lor~ t_c(\va{x},\va{z}) < t_{c,threshold})\\
    &\land~ rank(H_{\va{z_p}}) = n_{sv} ~\land~ t_c(\va{x},\va{z_p}) \geq t_{c,threshold}
    \end{split}\\
    \begin{split}\label{eq:se_normal}
    S_{SE} = \text{normal} ~\Leftrightarrow~ & \Phi_{serv} ~\land~ rank(H_{\va{z}}) = n_{sv} ~\land~ t_c(\va{x},\va{z}) \geq t_{c,threshold}\\
    &\land~ t_m(\va{x},\va{z}) \geq t_{m,threshold}
    \end{split}
\end{align}

Equations~\ref{eq:se_failed}\,-\,\ref{eq:se_normal} formalize the conditions for each state. $S_{SE}$ denotes the state of the state estimation service, $\Phi_{serv}$ the availability of the state estimation server, $\va{z}$ the set of field measurements, $\va{z_p}$ the set of field measurements augmented with pseudo-measurements, $\va{x}$ the set of state variables (state estimation results), $H(\va{z})$ and $H(\va{z_p})$ functions which relate $\va{z}$ and $\va{z_p}$, respectively, to $\va{x}$, and $t_c$ is the trust concerning data correctness issues. Note that this state classification considers a centralized state estimation. State classification of decentral and distributed architectures is part of future work. Since these architectures do not involve a central control room, the trust assessment becomes challenging because there is no central point where trust inputs can be gathered and trust values can be calculated. Moreover, field devices do not have advanced monitoring capabilities for computing trust. 

This section shows how trust can be used to measure the data correctness, which is then used to determine the operational states of the state estimation service. This corresponds to the second and third contributions of this paper (cf. Subsec.\,\ref{sec_contribution}).

\subsection{Operational States of Coordinated Voltage Control (CVC)}\label{subsec:voltagecontrol}

Disturbances in a power system, such as line failures and sudden changes in power generation, can impact the bus voltages. When a bus voltage exceeds a certain threshold (typically $\pm(5-10)\%$), voltage instabilities can arise~\cite{kundur2007power}. The voltage control service aims at remedying these voltage problems by suitably adjusting the output of controllable devices such as transformers or distributed energy resources. These devices are, henceforth, referred to as controllers. In this paper, a specific type of voltage control, namely \gls{cvc}, is considered. Typically, voltage control algorithms use local measurements at the bus to adjust the output of the corresponding controllers. In the \gls{cvc}, these measurements are transferred to a server, typically located in the control room~\cite{salih2015coordinated}. Here, optimization problems based on the received measurements are solved with the aim of adjusting the bus voltages, the results (or solution) of which are the coordinated adjustments of the various controllers. The existence of a feasible solution indicates that the underlying voltage problem can be remedied using the considered controllers. The corresponding control signals (known as setpoints) are then transferred back via the communication network to the controllers, which perform the required adjustments. These controllers can, however, also act based on local measurements in case the central control signals are absent. Unlike the state estimation service, which only provides monitoring information about the \gls{cpes}, the \gls{cvc} service also performs actuation based on the monitored data. 

A vital part of typical \gls{cvc} algorithms such as~\cite{salih2015coordinated,hird2004network,viawan2008coordinated} is the availability of reliable state estimation results as the \gls{cvc} optimization requires the knowledge of bus voltages in the grid. A failure of the state estimation service will result in a loss of grid observability for the \gls{cvc} service, thus hindering coordinated decision-making. In this case, the controllers do not receive setpoints from the control room and, therefore, can act only based on local measurements. When the state estimation service is in its limited state, i.e. when its results depend on pseudo-measurements, coordinated actions by the \gls{cvc} service are still possible but with an increased risk as the pseudo-measurements (based on historical data) may not capture certain recent events in the power system~\cite{narayan2021analyzing}. In this case, it can also be argued that the controllers should act based on local measurements rather than measurements from the control room. This is a design trade-off as coordinated actions, on the one hand, can remedy a broader range of disturbances but, on the other hand, rely on potentially incorrect state estimation results as well as the communication between the control room and controllers.  

To transmit the setpoints derived from the optimization results to the corresponding controllers, a communication network with an acceptable level of latency between the \gls{cvc} server and the controllers is required. Similar to the state estimation service, the acceptable latency depends on the design of the \gls{cvc} service as well as the severity of the disturbance, e.g., larger/faster voltage deviations require faster remedy~\cite{kundur2007power}. Furthermore, the ability of the \gls{cvc} service to remedy voltage problems in the \gls{cpes} is largely impacted by the trust in the state estimation results and the controllers. While the former is already covered in Subsection~\ref{subsec:stateestimation}, the latter is relevant considering the increasing risk of \gls{ict} disturbances in \glspl{cpes} such as hardware/ software failures and cyberattacks~\cite{narayan2019first}. In other words, a successful remedy of a voltage problem by the \gls{cvc} service requires controllers, which can be trusted and are reachable in time via the communication network. Based on these criteria, the state classification (cf. Section~\ref{sec_classification}), and the trust assessment (cf. Section~\ref{sec:trust}), the following operational states of the \gls{cvc} service reflecting its performance can be derived:

\begin{itemize}
    \item \textbf{Normal State}: The first requirement for the \gls{cvc} to be in normal state is that the state estimation service should not be in the failed state as the \gls{cvc} uses the state estimation results. Second, a feasible \gls{cvc} solution should exist with the controllers reachable on time and considering only trusted controllers. 
    

    \item \textbf{Limited State}: The \gls{cvc} is in the limited state if a feasible \gls{cvc} solution is only possible considering at least one untrusted controller. However, all involved controllers are still reachable on time. This state indicates potential problems when realizing the current \gls{cvc} solution due to the involvement of untrusted controllers. 
    
    \item \textbf{Failed State}: The \gls{cvc} is in the failed state if the state estimation service has failed, there is no feasible \gls{cvc} solution without using pseudo-measurements, or the required controllers are not reachable in time. In this state, coordinated actions are no longer possible and controllers must use local measurements. 
\end{itemize}

\begin{figure}[htb!]
    \centering
    \includegraphics[width=.75\textwidth]{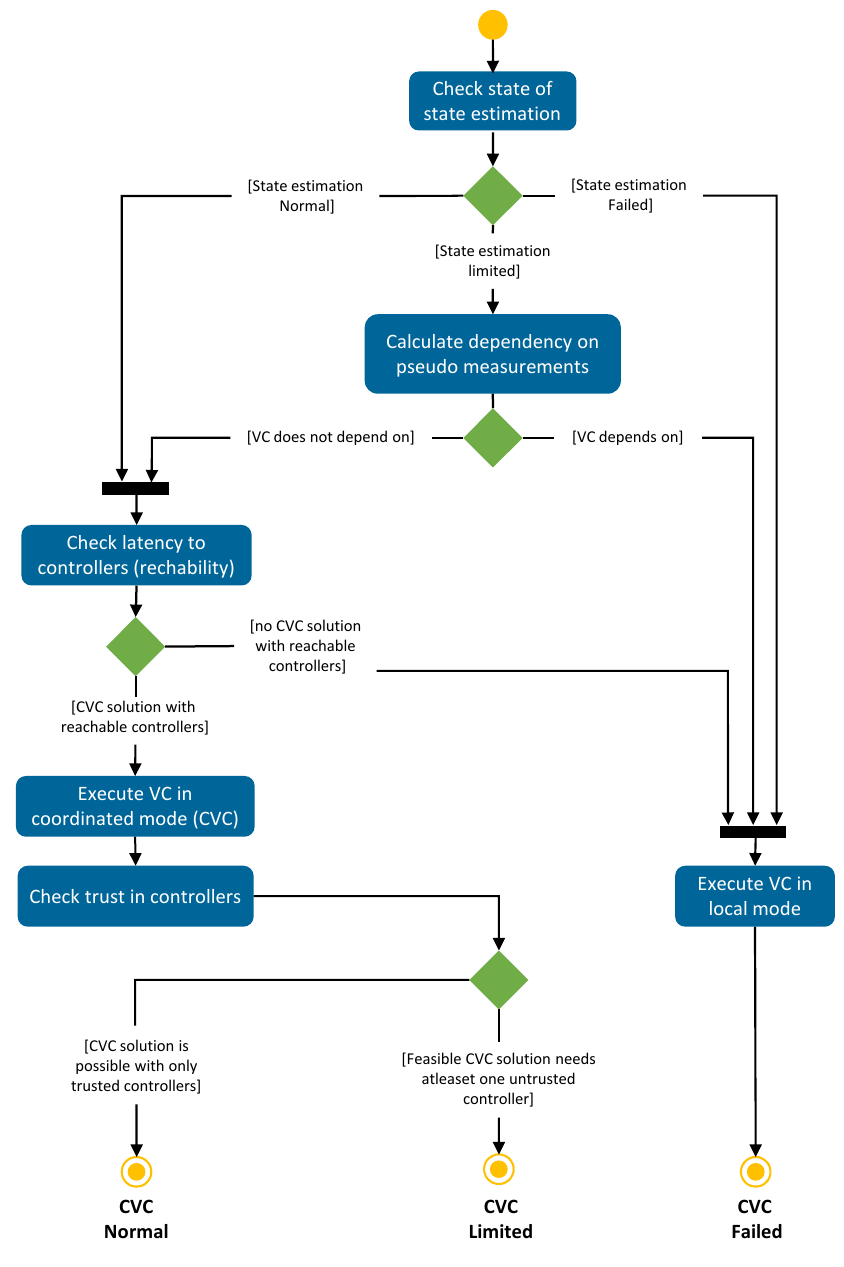}
    \caption{Activity Diagram showing states of Coordinated Voltage Control Service..}
    \label{fig:CVC_activity}
\end{figure}


Similar to the state estimation use case, appropriate actions from the operator are demanded in the limited state to improve the state, although the service has not failed. Note that the control room cannot directly monitor whether the controllers implement the received setpoints correctly or if the setpoints achieve the desired voltages at the power system buses. This can only be assessed in the next state estimation run using updated field measurements after the control action.

Figure~\ref{fig:CVC_activity} shows an activity diagram for the proposed process of deriving the state of the \gls{cvc} service. Since the \gls{cvc} service depends on the state estimation service, its state is checked first. A failure of the state estimation will cause the \gls{cvc} to fail, forcing the controllers to switch to the local mode. The same applies if state estimation is in the limited state and all possible \gls{cvc} solutions rely on pseudo-measurements. As in the case of state estimation, local voltage control is recommended in this case since pseudo-measurements are not as accurate as field measurements. As mentioned earlier, this is a design trade-off as \gls{cvc} can remedy a large range of disturbances (due to broader observability based on state estimation) but could also result in undesired control action due to a dependency on pseudo-measurements. Contrarily, local measurements can be trusted more but can only indicate local voltage problems. If the state estimation service has not failed and there exists a feasible \gls{cvc} solution that does not rely on pseudo-measurements, the reachability of the controllers, i.e. the ability to respond in time, is checked. A feasible \gls{cvc} solution that relies only on reachable controllers is then chosen. If no solution exists for which all controllers are reachable, the \gls{cvc} fails and the voltage control is executed in local mode. Otherwise, the trust in terms of data correctness of the reachable controllers is checked, which decides whether the \gls{cvc} is in the limited or normal state. Similar to the case of pseudo-measurements in the state estimation service, a design-specific threshold can exist to decide the maximum number of untrusted controllers that can be considered in a particular \gls{cvc} solution. If the number of untrusted controllers is higher than this threshold, the \gls{cvc} service is regarded as failed. Note that the \gls{cvc} optimization problem may have several feasible solutions, possibly ranging from best to marginally acceptable in terms of its quality. Since trust is checked only after calculating the \gls{cvc} solutions, it could happen that the best solution has a lower trust compared to a worse (but still feasible) solution. In this case, a trade-off has to be made between the quality of the solution and the trust in the controllers.
\begin{align}
    \begin{split}\label{eq:cvc_fail}
        S_{CVC} = \text{failed}~\Leftrightarrow~&(S_{SE} = \text{failed}) ~\lor~ \\ &(\forall y(\va{x}) \in Y ~|~ y(\va{x}) \stackrel{d}{\sim} \va{p}~\lor~ \exists a \in A_y~|~l_{serv,a} > l_{threshold})
    \end{split}
\end{align}
\begin{align}
    \begin{split}\label{eq:cvc_sev}
        S_{CVC} = \text{limited}~\Leftrightarrow~&(S_{SE} \neq \text{failed}) ~\land~ (\exists y(\va{x}) \in Y ~| ~(!(y(\va{x}) \stackrel{d}{\sim} \va{p}))\\
        &  ~\land~ (\forall a \in A_y~|~l_{serv,a} \leq l_{threshold}) \\ & ~\land~ (\exists a \in A_y ~|~ t_{a} < t_{threshold})
    \end{split}\\
    \begin{split}\label{eq:cvc_norm}
        S_{CVC} = \text{normal}~\Leftrightarrow~&(S_{SE} \neq \text{failed})\\ 
        & ~\land~ (\forall a \in A_y~|~l_{serv,a} \leq l_{threshold}~\land t_{a} \geq t_{threshold})
    \end{split}
\end{align}

Equations~\ref{eq:cvc_fail}\,-\,\ref{eq:cvc_norm} formalize the conditions for each state. $S_{CVC}$ denotes the state of the \gls{cvc} service, $y(\va{x}) = \{(v, a)\}$ denote a feasible \gls{cvc} solution that is a set consisting of pairs of voltage setpoints $v$ and the corresponding controller $a \in A$, $A_y$ be the set of controller which are involved in a CVC solution $y$, $Y = \{y(\va{x})\}$ be the set of all feasible \gls{cvc} solutions, $a \stackrel{d}{\sim} b$ a dependency of $a$ on $b$, $l_{serv,a}$ the latency between the \gls{cvc} server (or control room) and controller $a$, and $t_a$ is the trust of controller $a$ (data correctness cluster). 

This section shows how trust can be used to measure the data correctness, which is then used to determine the operational states of the CVC service. Similar to the state estimation service in Sec.\,\ref{subsec:stateestimation}, this corresponds to the second and third contributions of this paper (cf. Subsec.\,\ref{sec_contribution}).

\section{Discussion and Future Work}\label{subsec:discussion}

This section discusses the presented approach, open points and limitations. Subsection~\ref{subsec:discussion_alternatives} discusses potential adaptations for the state classification. Subsection~\ref{subsec:discussion_controllers} discusses the capabilities and limitations of controllers in the field to assess trust, while potential measures for system operators based on the state classification are presented in Subsection~\ref{subsec:discussion_operators}. Finally, Subsection~\ref{subsec:cpes_trust_discussion} discusses the application of trust in \glspl{cpes} from a holistic perspsective.

\subsection{Adaptions of State Classification}\label{subsec:discussion_alternatives}
As mentioned in Section~\ref{sec:use_cases}, the presented state classification for the state estimation and \gls{cvc} grid services is just one of many possibilities because grid services in \glspl{cpes} can vary drastically in their designs. One differentiator is the number of sensors in the power system.
While electric transmission grids are typically equipped with redundant sensors at nearly every power system bus, electric distribution grids are typically equipped with sensors only at specific buses without any redundancy. This is due to cost and a lower level of digitalisation in distribution grids~\cite{brunekreeft2015germany}. For the state classification of the state estimation service (cf. Figure~\ref{fig:SE_activity}), an alternative approach can be identified for grids with redundant measurements. Here, only trusted measurements could be considered by the state estimation algorithm, while untrusted field measurements could be discarded. This could potentially result in better state estimation results. However, discarding measurements in the case without redundant measurement (e.g., distribution grids) is impractical as it would directly violate the solvability condition (cf. Subsection~\ref{subsec:stateestimation}). However, the proposed state classification in Figure~\ref{fig:SE_activity} would be applicable to both cases.


Another point for discussion is the thresholds for availability, timeliness, and trust, which define state transitions. The diverse nature of grid services and their designs makes it challenging to identify universal thresholds. For instance, the timeliness requirement of the state estimation service can vary depending on the type of sensors, e.g., $\leq 0.1s$ as in~\cite{Kuzlu2014} and $\leq 1s$ as in~\cite{Kansal2012}. The same is true for the \gls{cvc} service, where a range of latency requirements from $1s$ to $30s$ can be identified~\cite{chenine2009survey}. However, the presented state classification does not provide concrete thresholds and can, therefore, be applied to any design by considering the corresponding thresholds. 

The three operational states of ICT-enabled grid services are established in the literature (cf. \cite{narayan2021analyzing}, \cite{haack2022hybrid}, \cite{klaes2020state}) and are similar to the widely-used operational states of the power system (cf. Fig.\ref{fig:example_states}). However, a more fine-granular state classification is required to assess the current performance quantitatively. Such a model can determine the current state and the \textit{depth of degradation}, e.g., a limited state near the border to the normal state is less critical than a limited state near the border to the failed state. As a first step, the limited state, which encompasses everything between full functionality and complete failure, could be elaborated. Since these states are intended for situational awareness, simplicity (ease of use and understanding) and complexity (having more states to capture the actual behaviour) should be balanced. 

\subsection{Abilities and Limitations for Controllers}\label{subsec:discussion_controllers}
In the \gls{cvc} use case, an important question is how controllers in the field can assess the trust in commands or setpoints from the control room. Field controllers, unlike a control room, do not have advanced monitoring capabilities, such as an \gls{ids}. This is relevant in case of, e.g., spoofing attacks, where attackers can masquerade as the control room and send malicious setpoints to controllers. In this regard, measures such as local plausibilisation or reputation are required. In a local plausibilisation, field controllers verify the received commands based on their intelligence. This is discussed in \cite{acatech2021} and is still a nascent area of research. Reputation refers to a trust assessment based on the experience of other controllers with commands from the control room~\cite{Steghofer2010}.

From a control room perspective, the state assessment of closed-loop services (with monitoring and control) like \gls{cvc} is more challenging than open-loop services (only monitoring) like state estimation. The performance of a closed-loop service can only be assessed when the impact on the power system is reported to the control room. This is typically only done in the next cycle of data acquisition and state estimation.

\subsection{Measures for System Operators}\label{subsec:discussion_operators}
One of the main benefits of the proposed state classification is that it provides the operator with information regarding the performance of \gls{ict}-enabled grid services, based on which suitable actions can be taken (in the case of state degradation). Specific proactive measures can also be taken to improve the resilience and robustness of the system, which can either prevent state degradation or improve the performance in the degraded state. Some examples are discussed in the following.

It can be seen in Figure~\ref{fig:SE_activity} that the state estimation service can fail either due to a server failure or loss of field measurements. In the case of the former, suitable actions include fixing the server with active backup~\cite{narayan2019first} or redeployment and reallocation of the state estimation algorithm using virtualization \cite{kruger2020real} can be performed. In the latter's case, actions such as repairing sensors and rerouting network traffic can be performed. These actions aim to mitigate the impact of the disturbance and improve the state of the grid service. 

Although the operational states only provide abstracted information about the performance of the grid service, a closer examination of the classification will provide details on which of the three properties, i.e., availability, timeliness, and data correctness (trust), are currently impacted. Regarding availability and timeliness issues in the state estimation service, pseudo-measurements can be used to prevent further state degradation. Here, increasing the number of sensors to have more redundant field measurements will make the system more robust to sensor failures, whereas measures to improve the quality of pseudo-measurements will ensure better performance in case of state degradation. Communication network measures such as packet prioritization, network function virtualization, and traffic shaping could facilitate fast data transfer and reduce network downtime in the case of disturbances~\cite{Fra2022}. The robustness of the communication network could also be improved by considering redundant paths and meshing during the design phase. These measures may vary depending on the impacted facets. In the case of a security issue (in particular, integrity), closing vulnerabilities of the affected components is a suitable measure. For reduced functional correctness based on, for example, excessive resource usage, the resources and the implementation should be checked. In this regard, the use of the trust facets, along with other tools, to trace the issue to its source can be investigated as part of future work. 

\subsection{Trust in  CPESs}\label{subsec:cpes_trust_discussion}
This paper presents a specific application of trust, i.e., to estimate the trust in components and data. The underlying trust approach can also potentially be used to gain a more holistic perspective of the health and state of a \gls{cpes}, which considers not only the electrotechnical measurements (e.g., voltages, currents) but also the performance of the ICT components and the grid services.

\begin{figure}[htb]
    \centering
    \includegraphics[width=\textwidth]{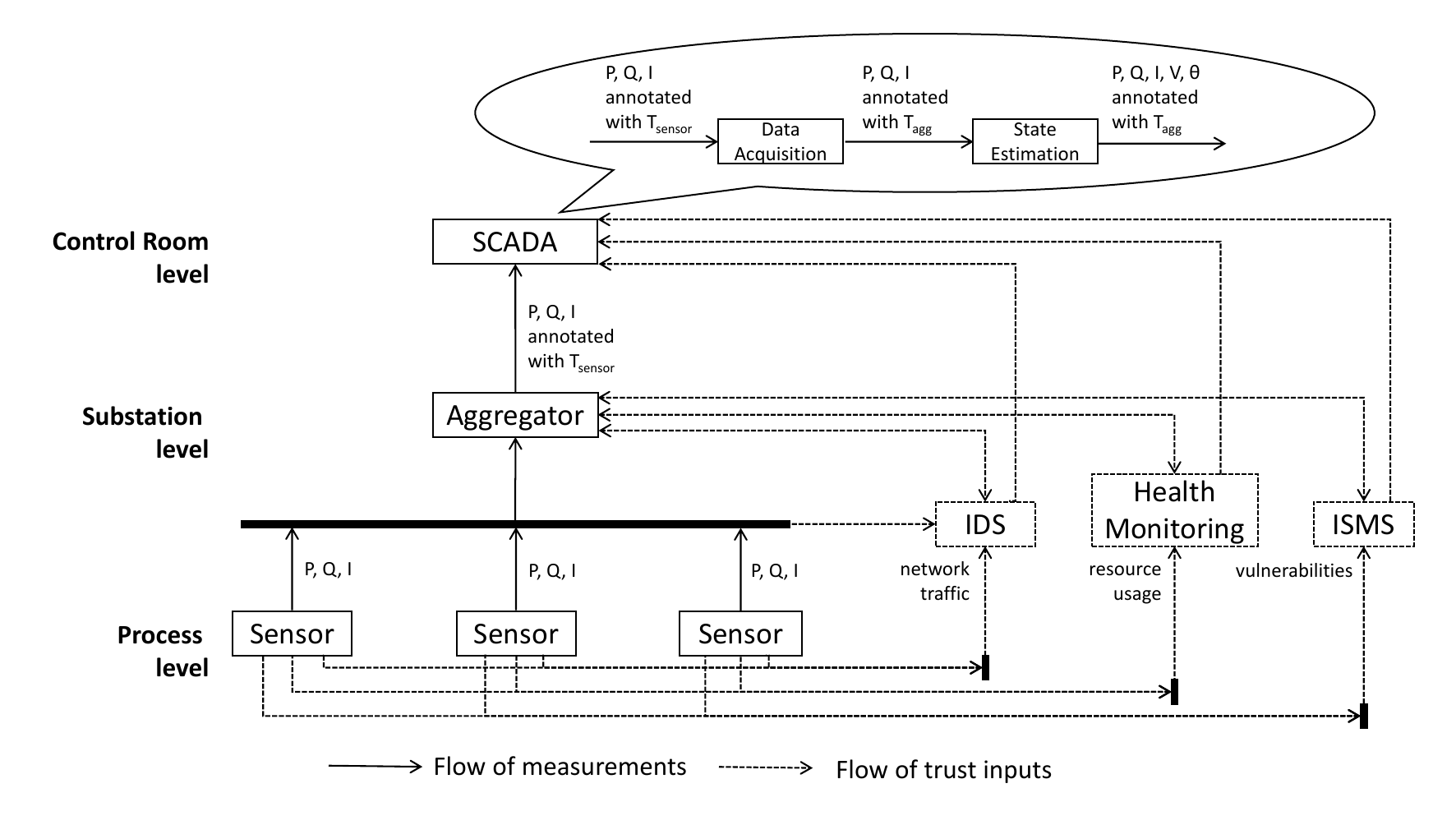}
    \caption{Exemplary System for Hierarchical Trust Assessment}
    \label{fig:trust_in_cpes}
\end{figure}

Because of the hierarchical structure of power systems, it may be feasible to also do the trust assessment hierarchically instead of centrally. Figure~\ref{fig:trust_in_cpes} shows an exemplary hierarchical trust assessment. The data flow for the electrotechnical measurements can be separated into three levels. On the process level, the sensors provide measurements such as active ($P$) and reactive powers ($Q$) and currents ($I$). These measurements are then sent to an aggregator located at the substation level. This corresponds to the power system buses in Fig.~\ref{fig:example_grid}. The aggregator collects the measurements from all the sensors and sends the packed measurements to a \gls{scada} system at the control room level. The first trust assessment for the sensors and their measurements can already be done at the aggregator using the information from an \gls{ids}, an \gls{isms}, and a health monitoring system, exemplarily. Therefore, the data transmitted to the \gls{scada} system would be enriched with a multivariate trust value. In the \gls{scada} system, a second trust assessment can be performed considering the trust in the aggregator. In Figure~\ref{fig:trust_in_cpes}, this is done based on trust inputs from the same trust sources, which also monitor the sensors. The output of the trust assessment, as shown in Figure~\ref{fig:psnatrust}, are state variables annotated with multivariate trust values, similar to the state estimation use case in Subsection~\ref{subsec:stateestimation}.

The future work regarding trust in \glspl{cpes} is three-fold. First, trust propagation in a hierarchical system with interdependent services must be modelled and analyzed. Second, subsystems that are not under the control of the system operator (e.g., third-party assets like solar panels in households) also need to be considered in the overall trust assessment, especially since \glspl{cpes} are organic systems with autonomous subsystems interacting with each other. The challenge here is that third-party assets may not have direct monitoring. Here, \textit{experience} (as in \gls{oc}-Trust) and possibilities to indicate own trustworthiness (trusted computing) are of interest. This contributes to a holistic trust assessment for \glspl{cpes} to determine the state and health of the overall system of systems.

\section{Conclusion}\label{sec:conclusion}

This paper applies the concept of \gls{psnatrust}, which is based on OC-Trust, to the operational state classification of ICT-enabled grid services. The operational states of grid services, based on literature, can be assessed based on three properties -- availability, timeliness, and data correctness. Unlike the first two properties, data correctness is challenging to measure due to the lack of a ground truth. Furthermore, the rising impact of cyber threats in \Glspl{cpes} increases the relevance of the data correctness property. The first contribution of the paper is the interpretation of \gls{psnatrust} and its facets to estimate data correctness considering various ICT disturbances. This is then applied to the operational state classification of ICT-enabled grid services, particularly for the data correctness property, which is the second contribution of the paper. As a third contribution, the operational states of two grid services, namely state estimation and \gls{cvc}, were discussed as use cases. In this regard, several possible future steps were also identified and discussed. The results provide the system operator (or the control room) with an enhanced situational awareness, particularly regarding the ICT-enabled grid services. In the case of state degradation, based on the state, alternative grid services could be used in addition to measures for improving the state.



\begin{acks}
This research has been funded by German Federal Ministry for Economic Affairs and Energy (BMWi) under agreement no. 03EI1020E (Resilienz-Monitoring für die Digitalisierung der Energiewende) and by Deutsche Forschungsgemeinschaft (DFG) – project number 359778999, as part of the priority programme DFG SPP 1984.
\end{acks}


\bibliographystyle{ACM-Reference-Format}
\bibliography{main}

\appendix

\end{document}